\documentclass[preprint, showpacs,amsmath,amssymb]{revtex4}
\usepackage{amssymb}
\usepackage{amsmath} %

\usepackage{graphicx}
\usepackage{amsfonts}
\usepackage{dcolumn}
\usepackage{bm}
\usepackage{multirow}

\begin{document}

\title{Coarse-graining the calcium dynamics on a stochastic reaction-diffusion lattice model}

\author{Chuansheng Shen$^1$}\email{schuansheng@aqtc.edu.cn}

\author{Hanshuang Chen$^2$}

\affiliation{$^1$Department of Physics, Anqing Normal University, Anqing, 246011, China \\
 $^2$School of Physics and Material Science, Anhui University, Hefei, 230039, China}


\begin{abstract}

We develop a coarse grained (CG) approach for efficiently simulating
calcium dynamics in the endoplasmic reticulum membrane based on a
fine stochastic lattice gas model. By grouping neighboring
microscopic sites together into CG cells and deriving CG reaction
rates using local mean field approximation, we perform CG kinetic
Monte Carlo (kMC) simulations and find the results of CG-kMC
simulations are in excellent agreement with that of the microscopic
ones.
 Strikingly, there is an appropriate range of coarse proportion $m$,
 corresponding to the minimal deviation of the phase transition point compared to
 the microscopic one. For fixed $m$, the critical point increases monotonously
 as the system size increases, especially, there exists scaling law between the deviations of the phase
 transition point and the system size.
 Moreover, the CG approach provides significantly faster Monte Carlo simulations which
 are easy to implement and are directly related to the microscopics,
 so that one can study the system size effects at the cost of reasonable computational time.

\end{abstract}
\pacs{87.16.A-, 05.50.+q, 82.20.-w, 05.10.-a} \maketitle

\section{Introduction}

 As a second messenger in living cells, calcium ions ($Ca^{2+}$) play a vital role
 in providing the intracellular signaling. Many important cellular processes
 and biological function, such as muscle contraction and synaptic transmission,
 are regulated by $Ca^{2+}$ signals
 \cite{NAT98000645,FAJ96001505,JBC95029644,AJP0000C235}.
$Ca^{2+}$ release is an inherently multi-scale problem, for
instance, in cardiac myocytes, the majority of the control of
calcium-induced-calcium-release (CICR) \cite{BPJ92000497,Ber2001215}
happens in the microdomain of the so-called diadic cleft, this
microdomain is  between the L-type voltage-gated $Ca^{2+}$ channels
and the ryanodine receptors. The ryanodine receptors \lq sense\rq \,
local
 [$Ca^{2+}$] in the diadic cleft positioned between the t-tubules and the
sarcoplasmic reticulum. The length scale of aforementioned
occurrences is on the order of nanometers and relevant time scales
range from micro- to milliseconds \cite{ANY06000362}. However, each
cell contains approximately 10,000 diadic spaces which act
independently \cite{PBMB06000136}. Typically, one is interested in
$Ca^{2+}$ currents at the whole cell level and higher. This is a
multi-scale problem. Given the limits of computational power, hardly
can we model an entire cellular cytoplasm by incorporating detailed
structural information.

 Some multi-scale models of CICR have been developed that successfully reproduced experimental
observations, as well as save computation largely
\cite{ANY06000362,PBMB06000136,JTB07000623}. However, these models
are based upon deterministic coupled ordinary differential equations
derived from biophysical mechanisms \cite{BPJ04003723}, and lack
accurate description of microscopic dynamics of calcium ion
channels. In fact, fluctuations are always exist in ion channels and
play a crucial role in $Ca^{2+}$ release mechanism
\cite{PRL00005664,PNAS03000506}.
 Recently, Vlachos and coworkers proposed a multiscale approach for coarse
 graining stochastic processes and associated Monte Carlo (MC) simulations
in surface reaction systems
\cite{JCP03009412,PNAS03000782,JCP03000250}. The method is efficient
in describing much larger length scales than conventional MC
simulations while still incorporating microscopic details, and
resulting in significant computational savings.  An overview of the
method is given in \cite{JCA07000253}.

In the present work, the multiscale approach was applied to a
relatively simple stochastic reaction-diffusion lattice model for
calcium dynamics in the endoplasmic reticulum (ER) membrane,
proposed by Guisoni \cite{PRE05061910,PRE06061905}. We coarse grain
the model and processes, and derive the coarse-grained (CG) surface
diffusion transition probability rates. By numerical simulations, it
is found that the results of CG kinetic MC (kMC) simulations are in
excellent agreement with that of the microscopic ones corresponding
to the optimal coarse proportion. Secondly, we study the system size
effects by fixing the coarse proportion, and find the phase
transition point increases monotonously as the system size
increases. Especially, there exists a scaling law between the
deviations of the phase transition point and the system size.
Finally, we investigate CPU time and find the approach provides
significantly faster MC simulations which are easy to implement and
are directly related to the microscopic one.

\section{Coarse-graining the lattice model}

\begin{figure}
\centerline{\includegraphics*[width=0.5\columnwidth]{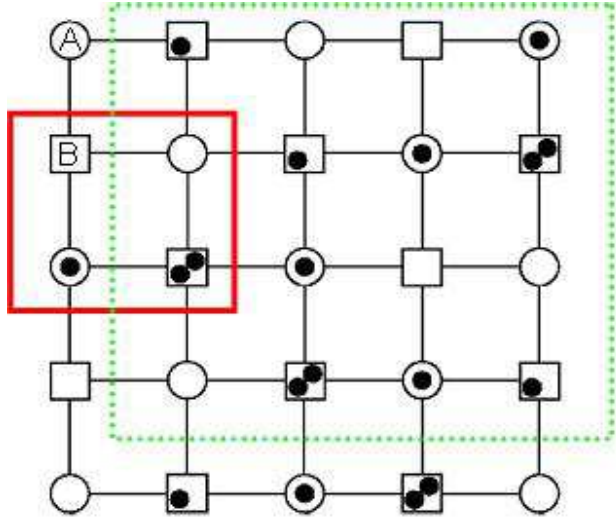}}
\caption{(Color online) Schematic illustration of coarse-graining
lattice model. Circle sites denote sublattice A and square cites
denote sublattice B. \label{fig1}}
\end{figure}

\noindent \emph{Microscopic Model\ } \textendash- We consider a
two-dimensional square lattice with two interpenetrating sublattices
A and B \cite{PRE05061910,PRE06061905} in ER membrane, as shown in
Fig. \ref{fig1}. Calcium channels are located only on the sites of
the sublattice B and calcium ions occupy not only the sites of the
sublattice A but also the sites of the sublattice B. A site $i$ of
the sublattice A can either be empty or occupied by at most one
calcium ion, the sublattice B take the values $0$, $1$ or $2$
corresponding to the closed, activating(open) and inhibiting state
respectively.

The dynamics of calcium ions in the model exhibit three stages. In
the first, spontaneous annihilation. If the site of the A sublattice
is occupied then it becomes empty with probability $q = (1 - p)a$ ,
here, $p$, related to the diffusion probability, $a$, related to the
annihilation process. In the second, diffusion. One of the four
nearest neighbor of site of the A sublattice, say site of sublattice
B, is chosen at random. A calcium ion then hops from a site of one
sublattice to a site of the other sublattice with probability $p$.
In the third, Catalytic creation. One of the four nearest neighbor
of site of the A sublattice, say site $j$ of sublattice B, is chosen
at random. If calcium channel $j$ is open then a calcium ion is
created at site $i$ with probability $r = (1 - p)(1 - a)$.

\noindent \emph{Coarse-Graining process} \ \textendash-
 In the paper, neighboring microscopic sites $q\times q$ are grouped together
into a CG cell, one can obtain a CG-lattice model with coarse cells.
Fig. \ref{fig1} shows an example of a coarse-graining lattice model
with $q =2$ and $q=4$, denoted by solid square and dotted square
respectively, here, $q$ is even because of two kind of sublattice A
and B.

 We define CG variables
\begin{equation}
\\
\tilde \eta  = \sum\nolimits_{i \in D_k } {s_i },
\\
\tilde \sigma _1  = \sum\nolimits_{j \in D_k } {\delta (x_j  - 1)}
,\tilde  \sigma _2  = \sum\nolimits_{j \in D_k } {\delta (x_j  - 2)}
\ \label{eq1}
\end{equation}
Here, microscopic variable $s_i $ and $ x_j$ denote the number of
calcium ions at the A and B respectively, and satisfies the
constraint
 $0 < \tilde \eta  < q^2 /2$,  $ 0 < \tilde \sigma _1 < q^2 /2$,  $0 < \tilde \sigma _2 < q^2 /2. $
since each coarse cell contains $q^2$ microcells. Equivalently we
may also consider the averaged version (termed below as coverage),
$s=2\tilde s/q^2, (s= \eta,  \sigma _1, \sigma _2)$. The dynamics of
calcium ions on the CG model has also three processes: a.
Spontaneous annihilation. b. Diffusion. c. Catalytic creation. The
table \ref{tab1} gives summary of processes and transition
probability rates for CG-kMC.

\begin{table}[htbp]
\caption{The processes and transition probability rates for CG-kMC.}
\label{tab1}
\begin{center}
\begin{tabular*}{15cm}{@{\extracolsep{\fill}}lllll}
 \hline\hline Process   & \multicolumn{3}{l}{Change of coarse variables}&Coarse transition probability rate\\
 & $\Delta \tilde \eta $ & $\Delta \tilde \sigma _1 $ & $\Delta \tilde \sigma _2 $ & \\ \hline
Annihilation & $ - 1$ & $\ \ 0$ & $\ \ 0$ & $ \tilde
W^1=a(1-p)\tilde\eta$\\\hline
Diffusion & $ + 1$& $ - 1$ &$\ \ 0$& $ \tilde W^2  = pq^2 (1 - \eta) \sigma _1 /2 $ \\
          &$ - 1$ &$ + 1$ & $\ \ 0$& $ \tilde W^3  = p\tilde \eta (1 -  \sigma _1  -  \sigma _2 ) $ \\
    　　　&$ + 1$&$ + 1$& $ - 1$& $ \tilde W^4  = pq^2 (1 -  \eta ) \sigma _2 /2 $ \\
    　　　& $ - 1$ & $ - 1$ &$ + 1$& $ \tilde W^5  = p\tilde \eta  \sigma _1 $ \\
\hline Creation  &$ + 1$ &$\ \ 0$ & $\ \ 0$ & $ \tilde W^6  = (1 -
p)(1 - a)q^2(1- \eta )\sigma _1 /2 $ \\ \hline\hline \\
\end{tabular*}
\end{center}
\end{table}

\section{Results and discussion}

Given a microscopic initial condition at random, following the
aforementioned rules in table \ref{tab1}, the coarse-grained calcium
dynamics is computed with periodic boundary conditions. But it needs
to make the computational demand of CG-kMC simulations per event the
same as that of microscopic MC ones.

\begin{figure}[h]
\centerline{\includegraphics*[width=0.6\columnwidth]{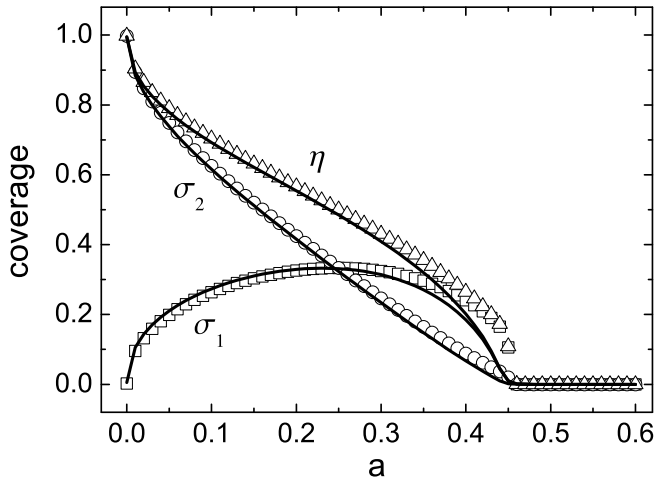}}
\caption{CG-kMC simulation results for the coverage $\eta, \sigma
_1, \sigma _2$ as a function of the parameter $a$ for the case
$p=0.5$ and $N\times N=200\times 200$. The coverage vanish at the
critical point $a_c\simeq 0.47$. Lines indicate micro-simulations,
symbols corresponds to CG ones with $q=40$. \label{fig2}}
\end{figure}

We perform CG-kMC simulations and microscopic simulations on a
square lattice with $N\times N=200\times 200$ sites, and plot the
coverage $\eta, \sigma _1, \sigma _2$ as a function of the parameter
$a$ in Fig. \ref{fig2}, where, $\sigma _1$ denotes the density of
open channels on sublattice B, $\sigma _2$ denotes the density of
inhibited channels on sublattice B, and $\eta$ denotes calcium ions
on sublattice A. Firstly we notice that the coverage predicted from
the CG-kMC simulations is in reasonably agreement with that of the
microscopic MC ones. Excitedly, the CG-kMC predicts the phase
transition point is in good agreement with that of microscopic MC
simulations. Indeed, small quantitative differences near the
critical point also exist, probably due in part to the fluctuation,
but still relatively small. These findings validate the CG approach
works well in simulating calcium dynamics.

\begin{figure}
\centerline{\includegraphics*[width=0.55\columnwidth]{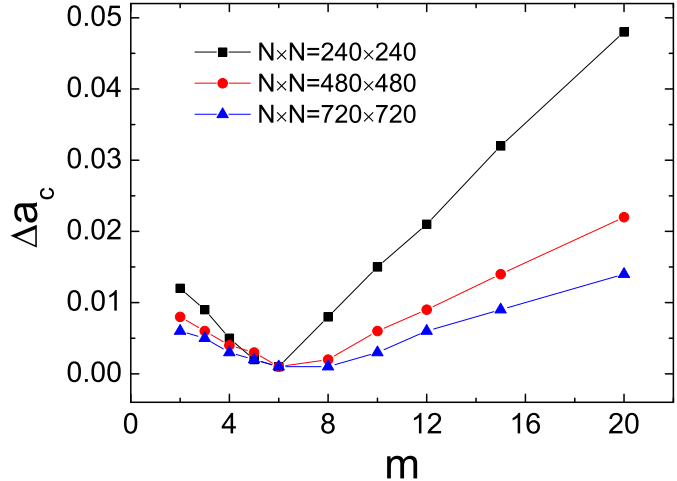}}
\caption{(Color online) The deviation $\Delta\alpha_c$ of the phase
transition points between CG-kMC and microscopic simulations vs the
coarse proportion $m$ for different system size $N\times N$.
\label{fig3}}
\end{figure}

To detect the phase transition point accurately, we need to choose
appropriate size $q\times q$ of coarse cell. But what a suitable $q$
is ? We define a coarse proportion $m$, $m = N/q$, namely, the
square root of the number of coarse cells. Changing $m$ we plot the
deviation $\Delta\alpha_c$ of the phase transition points between
CG-kMC and microscopic simulations for different system size
$N\times N$, as shown in Fig. \ref{fig3}. It can be seen that
$\Delta\alpha_c$ begin to decrease and then increase with the
increment of $m$, especially, the minimal $\Delta\alpha_c$ occurs
near the same point $m=6$ for different $N\times N$. When $N\times
N=720\times 720$, there are two values of $m$ corresponding to the
minimal deviation, seeming to a small plain appears. Furthermore,
the larger system size, the less deviation is. It is obvious that,
there exists an appropriate range of $m$ for coarse graining the
system precisely. Therefore we can fix $m$ and investigate the
effects of system size on the phase transition point.

\begin{figure}
\centerline{\includegraphics*[width=0.55 \columnwidth]{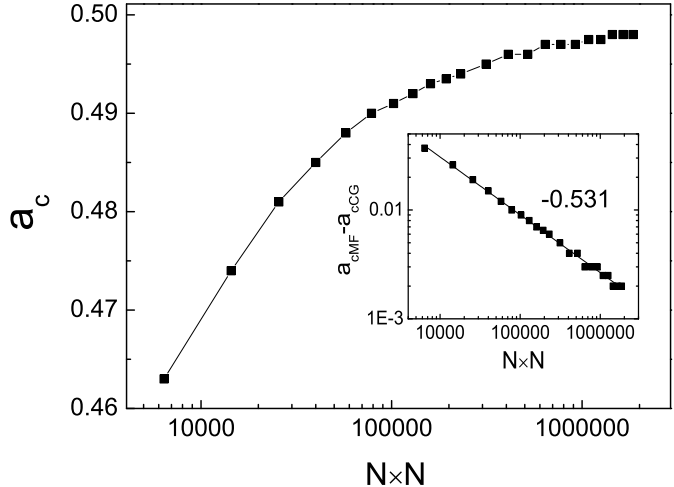}}
\caption{The dependence of the critical point $a_c$ on the system
size $N\times N$ for fixed $m\times m=6\times 6$. The inset gives
the scaling law of the deviations $a_{cMF}-a_{cCG}$ and $N\times N$,
where $a_{cMF}$ denotes the critical point of the MF model and
$a_{cCG}$ denotes the critical point of the CG model, the scaling
exponent is $-0.531$. \label{fig4}}
\end{figure}

For fixed $m=6$, we plot the critical point $\alpha_c$ as the
function of system size $N\times N$ in Fig. \ref{fig4}. Apparently,
$\alpha_c$ increases monotonously as $N\times N$ increases, and
approaches to $0.5$. Theoretically speaking, this asymptotic value
corresponds to the critical point of mean field (MF). The inset
gives the scaling relation of the deviations and the system size,
the scaling exponent is $-0.531$. To elucidate its accuracy, we have
also carried out coarse grained simulations with $m=8$ (not shown
here), and obtained a similar asymptote and power law. Therefore, we
can analyze the effects of the system size on the phase transition
point according to the scaling law and detect the critical point
accurately and rapidly.

\begin{figure}
\centerline{\includegraphics*[width=0.55 \columnwidth]{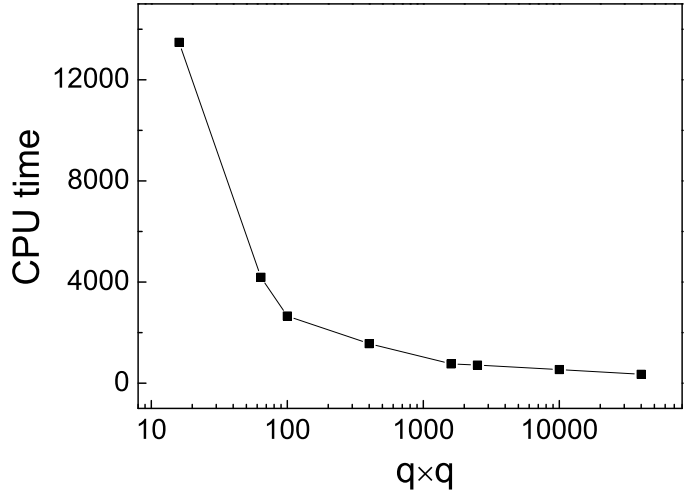}}
\caption{CPU time for CG-kMC as a function of coarse cell size
$q\times q$. \label{fig5}}
\end{figure}

 Finally, we exhibit the significant computational savings resulting
 from coarse-graining, as shown in Fig. \ref{fig5}. It can be seen that CPU time
 decreases monotonously as the size $q\times q$ of coarse cell increases.
 The larger size of coarse cell, the fewer CPU times. In this way,
 we can choose a bigger $q$ to save computational time. In fact,
 the computer time in kMC simulation with global update,
 i.e., searching the entire lattice to identify the chosen site,
 scales approximately as O($N^2$), but O($m^2$) in CG-kMC simulation.
 Accordingly, a $q$-fold reduction in the number of sites results
 in reduced computer time by a factor of 1/$q^2$. Therefore,
 coarse-graining can render MC simulation for the large scales feasible.

\section{Conclusion}

In this paper, we proposed an extensive CG model that can properly
describe calcium dynamics on ER membrane. By a great deal of
computer simulations, we demonstrated our model is highly effective
because the results of CG-kMC simulations are in very good agreement
with that of MC ones for a wide range of model parameters.
Interestingly, it was shown that there exists an appropriate range
of coarse proportion $m$, corresponding to the best estimation on
the phase transition point compared to the microscopic counterpart,
and such $m$ is almost insensitive to the change of the system size.
This make it possible to select an $m$ without beforehand unwanted
simulations for any real-world system size. Moreover, The CG-kMC
method provides significant reduction in CPU while retaining very
good accuracy in estimating the phase transition point. The larger
the level of coarse-graining $q\times q$ is, the larger
computational savings are, therefore we can obtain the phase
transition point quickly. Using the CG model, we also found that the
critical point increases monotonously as the system size increases.
Especially, there exists a scaling relation between the deviations
of the phase transition point and the system size.
A major advantage of the coarse model is that they have a direct
connection to the microscopic dynamics and can provide valuable
insights. Due to its reasonable accuracy and low computational
requirements, we anticipate that the methods outlined in this work
for simple systems will find widespread use in many realistic
systems.

\begin{acknowledgments}
This work was supported by the National Natural Science Foundation
of China (Grant No. 11205002). C.S.S. was also supported by the Key
Scientific Research Fund of Anhui Provincial Education Department
(Grant No.KJ2012A189).
\end{acknowledgments}

%
\bibliographystyle{apsrev}

\end{document}